\documentclass[preprint,12pt]{aastex}

\begin{document}

\title{X-ray Emitting Ejecta in Puppis A Observed with Suzaku}
\author{Una Hwang, Robert Petre, \& Kathryn A. Flanagan}

\begin{abstract}

We report the detection and localization of X-ray emitting ejecta in
the middle-aged Galactic supernova remnant Puppis A using five
observations with the Suzaku X-ray Imaging Spectrometer to survey the
eastern and middle portions of the remnant.  A roughly $3'\times 5'$,
double-peaked region in the north center is found to be highly
enriched in Si and other elements relative to the rest of the remnant.
The X-ray fitted abundances are otherwise well below the solar values.
While the ejecta-enhanced regions show some variation of relative
element abundances, there is little evidence for a very strong
enhancement of one element over the others in the imaged portion of
the remnant, except possibly for a region of O and Ne enhancement in
the remnant's south center.  There is no spatial correlation between
the compact [O III] emitting ejecta knots seen optically and the
abundance enhancements seen in X-rays, although they are located in
the same vicinity.  The map of fitted column density shows strong
variations across the remnant that echo earlier X-ray spectral
hardness maps.  The ionization age (as fitted for single temperature
models) is sharply higher in a ridge behind the northeast-east
boundary of the remnant, and is probably related to the strong
molecular cloud interaction along that boundary.  The temperature map,
by comparison, shows relatively weak variations.

\end{abstract}


\section{Introduction}

In recent years, X-ray spectral imaging has revealed X-ray emitting
ejecta even in a number of very old ($\sim$10,000 yr) remnants where
most of the X-ray emission is from shock-heated interstellar or
circumstellar matter (e.g, Hughes et al. 2003, Park et al. 2003,
Hendrick et al. 2003, 2005, Borkowski et al. 2006).  Puppis A is a
middle-aged supernova remnant whose X-ray emission is dominated by
shock-heated interstellar material.  It resides in a very complex
interstellar environment including large atomic and molecular clouds
(Dubner \& Arnal 1988, Reynoso et al. 1995) and a large-scale
interstellar density gradient (Petre et al. 1982).  It is, however,
one of a handful of remnants in which fast-moving ($\sim 1500$ km/s)
ejecta knots enriched in oxygen have been identified optically
(Winkler \& Kirshner 1985), and is thus a promising target for the
identification of hot, X-ray emitting ejecta.  The goal of the Suzaku
observations reported here is to search for and to localize X-ray
emitting ejecta in Puppis A.

The presence of O-enriched ejecta in Puppis A clearly indicates that
the progenitor was a massive star that synthesized the oxygen
hydrostatically, a conclusion that is further confirmed by the
presence of an X-ray emitting compact central source, believed to be a
neutron star (Petre et al. 1996, Winkler \& Petre 2007).  The
optically emitting oxygen knots are found in the central and eastern
regions of Puppis (Winkler \& Kirshner 1985).  More global oxgyen
ejecta enrichment had been indicated by high-spectral resolution X-ray
observations with the Focal Plane Crystal Spectrometer (FPCS) on the
Einstein Observatory (Canizares \& Winkler 1981).  The O enrichment
suggested by the Einstein FPCS could not be localized by it, however,
as the spectra were accumulated over rather large regions of the
remnant ($3'\times 30'$ and $6'$ diameter).  The identification and
localization of gas with enriched element abundances requires adequate
spectral imaging.  Einstein Solid-State Spectrometer observations with
a 6$'$ aperature did not find evidence for ejecta enrichment
(Szymkowiak 1985 PhD thesis), but a later survey by the ASCA
Observatory (Tamura 1995 PhD thesis) did produce evidence for
localized ejecta enrichment, mostly of Ne in the west and northwest.
Chandra observation of a small $8'$ square field at the eastern edge
of the remnant, where the shock is interacting with molecular and
interstellar clouds, hinted at small pockets of gas enriched with
ejecta (Hwang, Flanagan, \& Petre 2005).

In this paper, we present a partial X-ray imaging and spectral survey
of the Puppis A supernova remnant with the CCD detectors on the Suzaku
Observatory.  Five observations cover about two-thirds of this large
50$'$ diameter remnant to the east and center, including the locations
of the optically emitting O ejecta knots.  Compared to ASCA, Suzaku
has only slightly improved angular resolution, but much better
sensitivity and spectral resolution, particularly at low energies.
The Suzaku XIS also surpasses the Chandra CCDs for sensitivity and
spectral resolution at low energies, a feature that is particularly
important for studying the X-ray emitting O and Ne ejecta.

\section{Observations and Analysis}
 
Five observations of Puppis A were performed with the CCD-based X-ray
Imaging Spectrometer (XIS) at the beginning of the AO1 guest observing
phase of the Suzaku Observatory.  The XIS features four separate
imagers, one illuminated from the back (XIS1), the other three from
the front (XIS0, XIS2, subsequently defunct, and XIS3).  Details can
be found in Mitsuda et al. (2007).

The exposure times were optimized for observation of the oxygen
emission lines (O VII resonance at 574 eV, the dominant component of
the He-like O triplet in Puppis A, and O VIII Ly$\alpha$ at 653 eV;
see Winkler et al. 1981).  The observations, which are summarized in
Table 1, cover most of the bright eastern and central portions of the
remnant, but not the west.  Given the very soft spectrum of the source
and the overall high count rates of the Bright Eastern Knot (BEK),
Northeast (NE) and Interior (I) fields, data from the Hard X-ray
Detector on Suzaku were not telemetered, in favor of data from the
XIS.  Additionally for these three observations, 4s burst mode was
used for the FI chips in order to further reduce the telemetry load
(in this mode, the detector does not collect data for 4 s out of the
normal 8 s data cycle).  Shorter exposures also reduce the impact of
pulse pile-up in the detector, which is likely to be an issue at the
10\% level in the bright regions of the remnant.  The observations
were carried out without the use of charge injection to moderate the
effects of radiation damage on the CCDs.  We use the initial release
version 0.7 of the data, after performing the data screening
recommended in the Suzaku Data Reduction
Guide\footnote{http://heasarc.gsfc.nasa.gov/docs/suzaku/analysis/abc}.
\begin{deluxetable}{lccccc}
\tabletypesize{\footnotesize}
\tablecaption{AO1 Suzaku XIS Observations of Puppis A}
\tablewidth{0pt}
\tablehead{
\colhead{Field}&\colhead{Obs ID}&\colhead{Date}&\colhead{Coordinates}&\colhead{Exposure Time (ks)}&\colhead{FI Burst Mode}}
\startdata
Bright Eastern Knot (BEK) & 501086010 & 2006 Apr 17 & 125.9422, -42.9626 & 16.8 & 4s \\
Northeast (NE) & 501087010 & 2006 Apr 17-18 & 125.7579, -42.7332 & 20.7 & 4s \\
Interior (Int) & 501088010 & 2006 Apr 17 & 125.5915, -42.9172 & 20.3 & 4s \\
Southeast (SE) & 501089010 & 2006 Apr 18 & 126.0121, -43.2023 & 29.8 & 4s, none \\
South (S) & 501090010 & 2006 Apr 18-19 & 125.6826, -43.1664 & 31.2 & none \\
\enddata
\end{deluxetable}
\subsection{Imaging Analysis and Results} 

For all five fields, the full-CCD spectra are qualitatively similar.
They fall off rapidly above energies of 3-4 keV, with no evidence for
harder emission, and have numerous strong emission line features.
These are the He$\alpha$ blends of O (dominated by the resonance
line), Ne, Mg, Si, and S, plus weak Fe L features near 827 eV
corresponding to Fe XVII (cf. Winkler et al. 1981).  Figure 1 shows
the BI illuminated XIS1 spectrum for the South field as an example,
with the prominent emission lines and blends labelled. In this and all
other spectra considered here, the channels are binned to provide at
least 25 counts per channel.

\begin{figure}
\centerline{\includegraphics[scale=0.40,angle=-90]{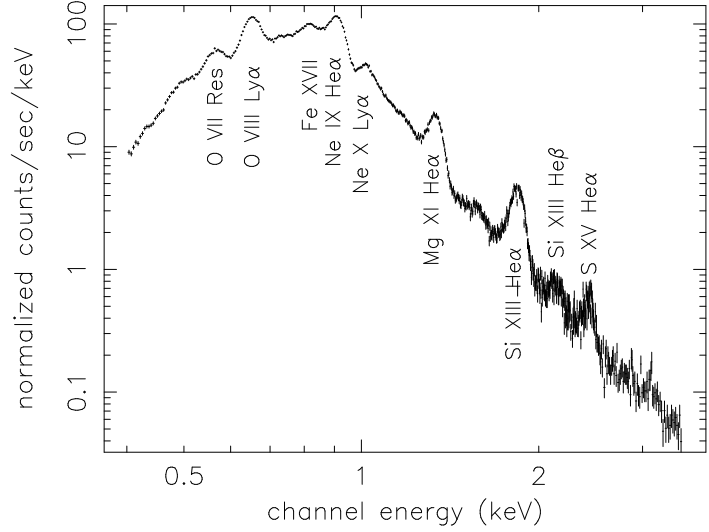}}
\figcaption{The XIS1 spectrum for the South field of Puppis A with 
the prominent emission lines and blends labelled.}
\end{figure}

\begin{deluxetable}{llll}
\tabletypesize{\footnotesize}
\tablecaption{Image Pulse Height and Energy Cuts}
\tablewidth{0pt}
\tablehead{
\colhead{Image}&\colhead{Pulse Heights}&\colhead{Energies (keV)$^*$}&\colhead{Total Counts ($10^4$)}}
\startdata
Cont 0            & 100-125 & 0.36-0.46 & 18\\
C He$\alpha$      & 125-145 & 0.46-0.53 & 51\\
O VII He$\alpha$ (Res) & 145-164 & 0.53-0.60 & 87\\
O VIII Ly$\alpha$ & 164-192 & 0.60-0.70 & 266\\
Cont 1            & 192-216 & 0.70-0.79 & 259\\
Fe L (Fe XVII)    & 216-236 & 0.79-0.86 & 313\\
Ne IX He$\alpha$  & 236-268 & 0.86-0.98 & 573\\
Ne X Ly$\alpha$   & 268-300 & 0.98-1.09 & 322\\
Cont 2            & 312-348 & 1.14-1.27 & 166\\
Mg XI He$\alpha$  & 348-390 & 1.27-1.42 & 175\\
Cont 3            & 390-420 & 1.42-1.53 & 42\\
Si XIII He$\alpha$& 475-538 & 1.74-1.96 & 63\\
Si XIII He$\beta$ & 563-633 & 2.05-2.31 & 15\\
S XV He$\alpha$   & 633-710 & 2.31-2.59 & 12\\
Cont 4            & 710-860 & 2.59-3.13 & 7.4\\
\enddata
\tablenotetext{*}{The energies are computed for a gain of 3.65 eV per pulse height channel.}
\end{deluxetable}

We produced energy-selected images of all the notable line features in
the integrated spectrum, as well as regions of continuum emission, as
summarized in Table 2.  We perform a simple correction for the
detector exposure by making the individual line images for each
detector for each observation, then trimming several pixels at all
four edges and dividing by the average exposure time; i.e., we do not
correct for vignetting.  No attempt was made to subtract background as
the images are all bright.  The final mosaicked images have 8.4$''$
pixels and are scaled to a fiducial exposure time, taken to be the
16.8 ks exposure time for observation 50108600.  The Si He$\beta$ and
S He$\alpha$ images have the fewest photons and so were binned by a
further factor of four in each dimension to improve signal-to-noise.

The combined mosaicked line images for all four XIS detectors are
shown in Figure 2; the images mosaicked separately for the BI and FI
detectors are not shown but are similar to those in the Figure.  The
continuum images are also not shown, but are similar to the line
images in general morphology.  Spatial variations in the line emission
are further highlighted by difference images, five of which are shown
in Figure 3.  These difference images are line images with the
broadband image normalized to it and subtracted.  While we refer to
the images in Figure 2 and 3 as ``line images'', they are in reality
simply energy-selected images that include any emission in the
relevant energy range, whether from line or continuum.  In Figure 2,
the O VII image is overlaid with surface brightness contours of the
ROSAT High Resolution Imager mosaic of the entire remnant (Petre et
al. 1996) and the J2000 coordinates.  Coordinates are also given for
the images in Figure 3 to facilitate identification of morphological
features discussed below.

The so-called Bright Eastern Knot (BEK), located just inside the
bottom of the straight eastern shock front, is prominent in all the
images.  Its brightness peaks at roughly (J2000) $\alpha$=126.04,
$\delta$=-42.97 (see Figures 2 and 3).  It is known to be a
complicated interaction between the supernova remnant shock and
multiple interstellar clouds (Petre et al. 1982, Hwang et al. 2005).
Given that the eastern part of the BEK has lower temperature and
ionization age conducive to stronger O line emission (Hwang et
al. 2005), it is not surprising that O emission is seen to be
relatively prominent at the easternmost edge of the BEK, and Ne and Fe
L emission is strong in a larger region extending towards the remnant
interior (Figure 3).  We also point out some of the other
morphological features that we will later discuss in more detail.
These include the elongated region extending southward from the BEK
region (SE Arm, roughly $\alpha$=126.05, $\delta$=-43.15), and the low
surface brightness region that is just west of it (SE Low SB).
Extending north by northwest from the BEK is the northeast shock
front (NE SF); seen by Suzaku, it is bright in the images of Ne, Mg,
Si, S, but not of O.
The relative prominence of lines of higher energy as opposed to oxygen
lines is consistent with the harder spectra seen interior to the NE SF
in ROSAT observations (Aschenbach et al. 1994, Hwang et al. 2005).
The remnant's interior is populated with additional patchy clumps of
emission.  The most striking of these is seen most prominently in the
Si images, just inside the northernmost apex of the remnant in the Si
images (it is also seen clearly in Ne, Mg, and S emission; we call
this the ``Si Knot'' ($\alpha$=125.70, $\delta$=-42.76).
Incidentally, the neutron star in Puppis A is visible in Si and S, and
in negative in the Ne IX difference image, at the lower right of the
image mosaics (i.e., in the lower center of the remnant at
$\alpha$=125.48, $\delta$=-43.01).

\begin{figure}
\centerline{\includegraphics[scale=0.8]{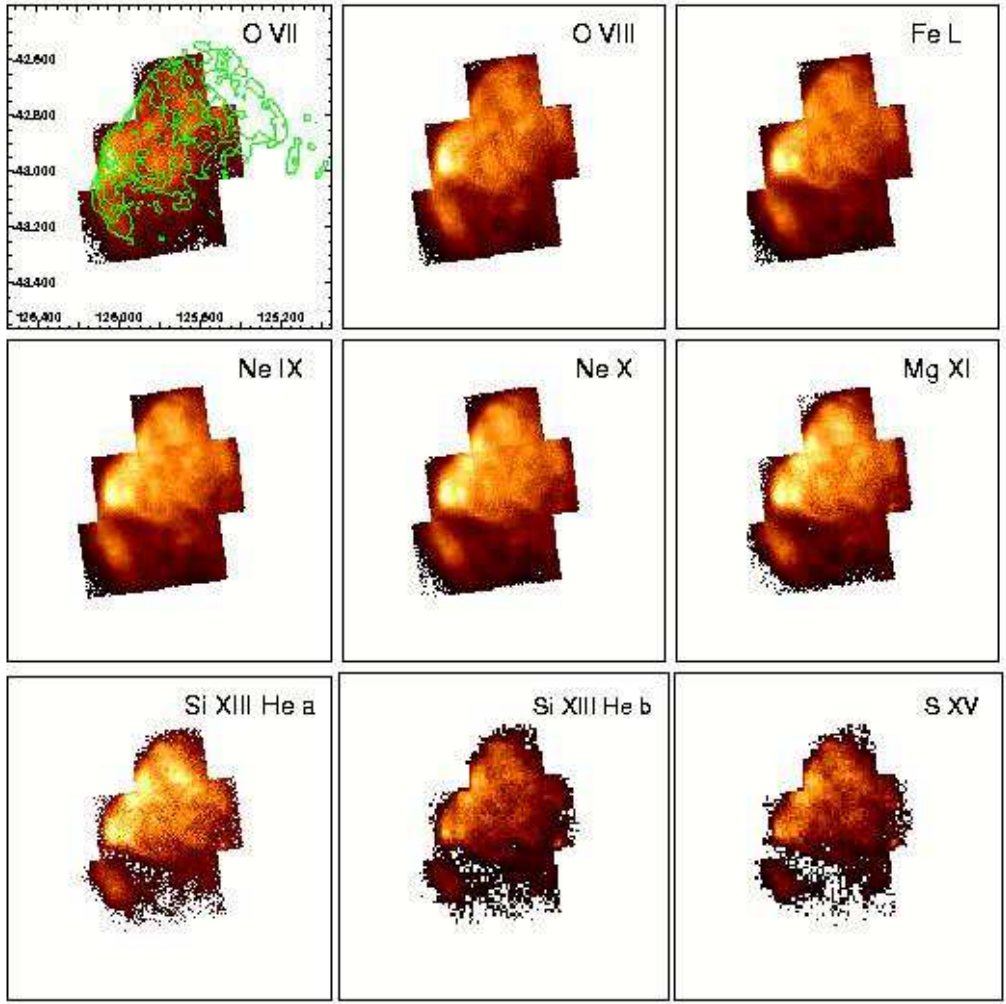}}
\vskip -1.5 in
\caption{Energy selected line image mosaics of all four XIS detectors
combined.  From left to right: (top) O VII Res, O VIII Ly$\alpha$, Fe
L (Fe XVII) (middle) Ne IX He$\alpha$, Ne X Ly$\alpha$, Mg XI
He$\alpha$ (bottom) Si XIII He$\alpha$, Si XIII He$\beta$, S XV
He$\alpha$.  The final two images have been binned further by a factor
of four to improve the signal-to-noise level.  The O VII image is
labelled with J2000 coordinates and overlaid with surface brightness
contours of the ROSAT HRI mosaic (Petre et al. 1996) to indicate the
full extent of the remnant.  The neutron star can be seen in the
images in the bottom row.}
\end{figure}

\begin{figure}
\centerline{\includegraphics[scale=0.8]{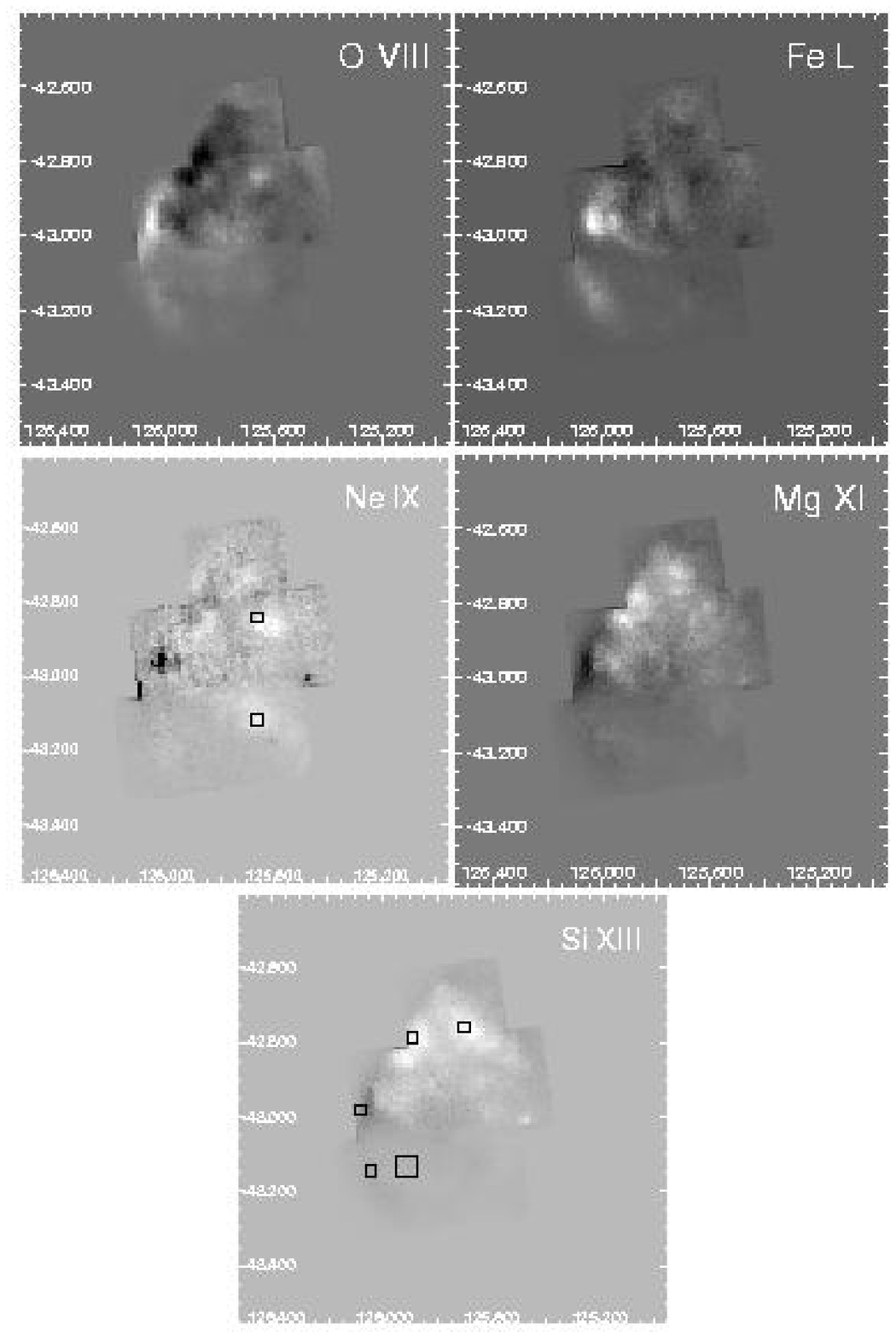}}
\vskip -1.5 in
\caption{Selected line images from Figure 2, with scaled broadband
image subtracted.  White indicates excess emission compared to
broadband; black indicates a deficit of emission.  For reference, the
spectral regions presented in Figure 6 are shown in the panels for Ne
IX (top: O Knot Center and bottom: Ne Knot South) and Si XIII
(counterclockwise from top: Si Knot North, NE SF, BEK, SE Arm and SE
Low SB)}
\end{figure}

Line-to-continuum ratio images (``line equivalent width'' or EQW) can
be used to make a quick assessment of whether enhancements in the
images are particularly due to enhanced line emission.  The continuum
underlying the line emission is first estimated (either by
interpolation or extrapolation from the continuum in nearby energy
regions,
or by extrapolating from a single adjacent continuum region),
then it is subtracted from the ``line'' image before taking the ratio
(the procedure is essentially that used by Hwang et al. 2000).  We
examined EQW images for O VIII Ly$\alpha$, Ne He$\alpha$, Mg
He$\alpha$ and Si XIII He$\alpha$ with particular reference to the
emission enhancements at the BEK, the northeastern shock front, and
the various interior knots.
None of the EQW images is bright at the BEK, which is in line with our
understanding that the BEK is associated with shocked interstellar
clouds.  The northeast shock front is also absent from the EQW images
for the same reason.  Of all the features seen in the EQW images, the
strongest by far is the roughly $3'\times 5'$ Si Knot seen in the Si,
Ne, Mg, and S images.  It completely dominates the Si EQW image which
is shown in Figure 4.  (This image is binned by a factor of four
relative to the Si He$\alpha$ line image in Figure 2.)  We do not show
the O, Ne, or Mg images here, in favor of the abundance maps from
spectral fitting to be discussed below, but note that O and Ne EQW
also appear high in a few regions, particularly in the southern
portion of the Si Knot and in the SE Low SB region adjacent to the SE
Arm.

\begin{figure}
\centerline{\includegraphics[scale=0.8]{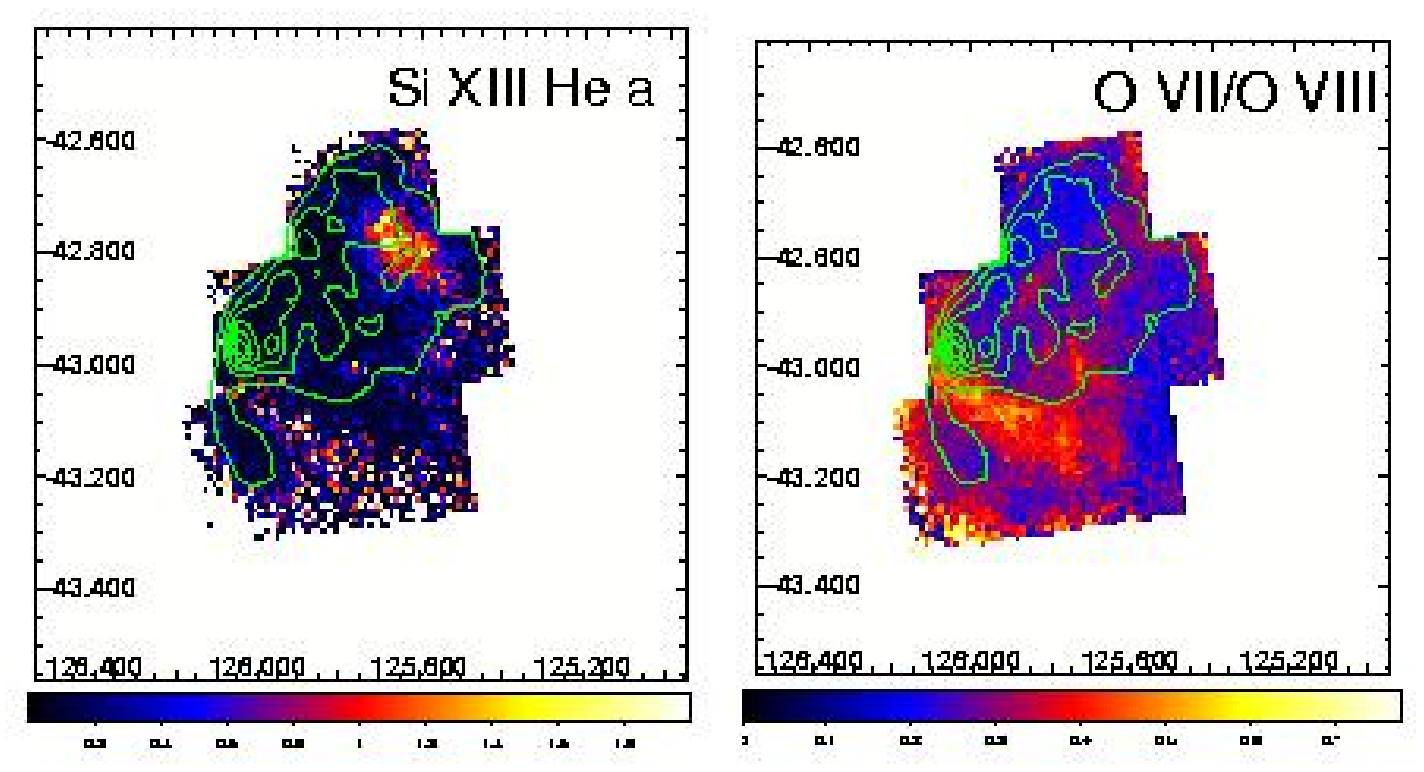}}
\vskip -2.5 in
\caption{Left: Line-to-continuum, or ``equivalent width'' image for
Si XIII He$\alpha$.
Right: Line ratio image for O VII Res/O VIII Ly$\alpha$.}
\end{figure}

Element abundances aside, we also briefly consider the O VII Res/O
VIII Ly$\alpha$ ratio image as an example of a single-element
line-ratio image that can be used to diagnose conditions in the X-ray
emitting plasma (Figure 4; the analogous image for Ne is similar and
not shown).  Ideally the ratio should be for line
emission only, but we have taken the ratio without subtracting the
continuum; the true continuum level is difficult to determine at this
spectral resolution, and the O VII line in particular is rather weak.
The image highlights the SE Low SB region that extends southward from
the BEK, as well as the eastern edge of the BEK, but cannot tell us to
what extent the O VII enhancement is caused by temperature, ionization
age, or column density differences.  Imaging spectral observations
with Chandra indicate that the BEK is composed of at least two
physical components, with the easternmost of these at lower
temperature, lower ionization, and higher column density (Hwang et
al. 2005).  Except for the higher column density, these would
contribute to the observed enhancement of O VII/O VIII at the eastern
edge of the BEK.  Likewise for the low surface brightness region,
further constraints from spectral analysis are needed to understand
the enhancement observed.

\section{Spectral Analysis}

While the EQW images provide an easily accessible overview of where in
the remnant the line emission is enhanced, a quantitative
interpretation is not possible from these images alone.  First, it
must be known how temperature, ionization, and element abundances
affect the line strengths relative to the continuum, and this depends
on the underlying spectral model assumed.  Moreover, the parameters of
any model are not expected to be uniform throughout the remnant.
Second, the continuum estimation can be accurate only if the
``continuum'' regions are truly line-free to start with, but this is
not usually a safe assumption for lines such as those of O or Ne that
lie at energies where the spectra are dense with line emission.
Finally, a two-point interpolation is the only feasible method to
estimate and subtract the continuum here, but it lacks sophistication
and probably accuracy.  EQW images are thus most likely to be useful
if the line emission is very strong and well-isolated; for example, in
tracing the qualitative large-scale morphology of known, strong ejecta
enhancements as in Cas A (Hwang et al. 2000).  They are less reliable
when searching for relatively modest ejecta enhancements.  To support
the conclusions of the imaging analysis, it is thus necessary to turn
to the X-ray spectra, as we do next.

We extracted spectra across the portion of the remnant imaged by
Suzaku in a grid following the procedure used by Hwang et al. (2005)
for Chandra observations of the BEK.  Spectra were extracted for 259
square and rectangular regions of angular size 100-200$''$, with each
spectrum containing at least 10,000 counts.  We focus on the spectral
results for the BI chip.

The spectra were binned to 25 counts per channel and were fitted using
XSPEC with nonequilibrium ionization (NEI) models for a plane parallel
shock, modified by interstellar absorption using cross sections from
Morrison \& McCammon (1983).  The models are characterized by the
temperature and the limiting ionization ages; we take the lower limit
for the ionization age to be fixed at $n_et = 0$ cm$^{-3}$~s and have
fitted only for the upper limit.  Element abundances were varied for
O, Ne, Mg, Si (with S, Ar, and Ca linked to Si), and Fe (with Ni
linked to Fe).

The spectral models must be converted to pulse-height spectra for
comparison to the data.  This conversion is carried out with an
effective area file giving the total efficiency to detect a photon of
a given energy, and a response matrix for the distribution of photon
energy with pulse height.  Both of these responses can vary depending
on the region of the detector used and the specific observation
conditions.  An important time-dependent effect is the reduction in
XIS low energy efficiency caused by the build-up of contaminants on
the optical blocking filter (Koyama et al. 2007).  This effect is
included in the effective area calculation (Ishisaki et al. 2007).
CCDs are also damaged by radiation with the passing of time causing
degradation in the energy resolution and gain.  While this effect is
now ameliorated by charge injection, charge injection was not used for
these early observations.  We compensate by fitting the spectrum with
an overall scaling factor for the gain (usually 0.993 to 0.996).

The observations were spread out over only two days, and there was no
discernable change in the response matrices computed for the various
observations.  We use the matrix appropriate for an observation using
the entire CCD chip because the current matrices do not reproduce
differences in response across the detector.

The effective areas are another story.  Ideally, the effective area
should be computed for the specific region on the detector used for a
given spectrum, but this requires time-consuming ray-tracing
simulations for each spectrum.  This was impractical for us given
limitations in computer power and the large number of spectral regions
involved.  The good signal-to-noise level of our data allowed us to
use relatively small spectral regions, but this makes the effective
area compuations even more time-consuming.  We therefore compared the
results of spectral fits using several ray-tracing simulations of the
effective area: five for relatively large regions ($3'$ radius) on the
detector (center, top and bottom to the left and to the right) and
also a smaller 2$'$ grid region.  Using the same spectrum and response
matrix, but different effective area files, we find that the
temperatures remain very consistent; abundances and ionization ages
vary from a few to 10 percent, and the column density variation by
less than twenty percent.  For all the fits presented in this paper we
used the simulation for a $3'$ radius region on the center of the
detector, which amounts to neglecting the energy- and
spatial-dependent effects of vignetting and the contaminant on the
optical blocking filter.  The vignetting removes photons from the edge
of the detector and is more severe at high energies, while the
contamination removes photons from the center of the detector and is
more severe at low energies (see Ishisaki et al. 2007).  The two
effects tend to compensate, but the overall effect is
complicated. Since the spectrum of Puppis A peaks at around 1 keV,
neither effect should be too severe.  Given that the variation in the
spectral parameters across the remnant is significantly larger than
the spread in spectral parameters that we find from using different
effective area computations, we consider this to be an acceptable
simplification.  The biggest limitation imposed by the lack of fully
correct effective area calculations throughout the remnant is that we
are not able to obtain accurate model emission measures.

Given the uncertainties in the spectral response described above, we
do not give numerical values of the spectral parameters obtained for
the entire grid.  Rather we present an overview of the fitted
temperatures, ionization ages, column densities, and abundances in
Figure 5, with the smoothed Suzaku broadband mosaic image contours
overlaid for reference.  The fitted parameters are represented for
each region by the color scale.  Spectra and the fitted model
parameters are shown in Figure 6 and Table 3 for a small sample of
interesting spectral regions.

\begin{figure}
\centerline{\includegraphics[scale=0.8]{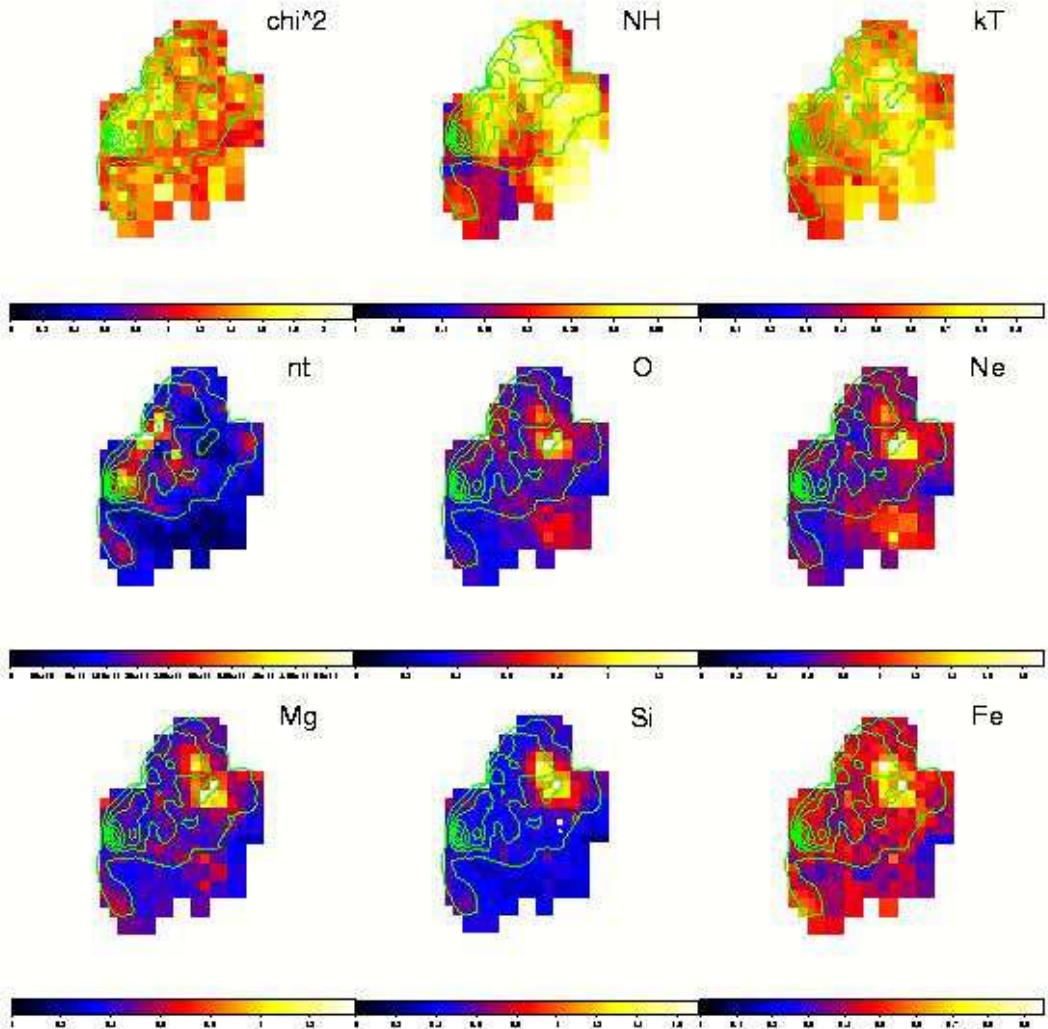}}
\vskip -1.5 in
\caption{Maps of the fitted parameters for plane-parallel shock
models, overlaid with smoothed contours of the Suzaku broadband image
mosaic: $\chi^2/d.o.f$, column density $N_H$ ($10^{22}$ cm$^{-2}$),
temperature kT (keV), ionization age $n_et$ (cm$^{-3}$ s), and element
abundances of O, Ne, Mg, Si, S, and Fe by number relative to
solar. The Si abundance map has the explosion center inferred by
Winkler et al. 1985 marked as ``E''.}
\end{figure}

Figure 5 shows that the quality of the fits is generally fair, with
all but a few values of the $\chi^2$ per degree of freedom falling
below 2.  While not formally acceptable in a statistical sense, we
consider it acceptable given that the model is simple and many of the
spectra a few tens of thousands of counts or more.  Column densities
are distributed mostly between about 0.13 to 0.35 $\times 10^{22}$
cm$^{-2}$, and are at the higher end of this range in the north and
center, away from BEK.  The regions of high absorption generally
correspond to the spectrally harder regions in the ROSAT images
(Aschenbach 1994).  The upper ionization age is typically
$0.5-2.5\times 10^{11}$ cm$^{-3}$s, but is particularly high along a
relatively narrow ridge behind the straight portion of the NE SF on to
the inside of the BEK.  Since the gas near the shock front should have
been shocked relatively recently, these high ionization ages (which
exceed $3\times 10^{11}$ cm$^{-3}$s) are consistent with high gas
densities.  The lowest ionization ages occur in the SE Low SB region,
as might be expected since the low surface brightness indicates low
gas densities.  The temperature map is somewhat patchy, but more
uniform than the others, with temperatures distributed fairly
uniformly between 0.5-0.8 keV.

The abundance maps all show a prominent double peak at the Si Knot in
the north central region.  Two strong peaks are seen in Mg, Si and Fe,
but the southernmost of these is more prominent in O and Ne.  All the
fitted element abundances have their peak values in this region; these
are all above solar, except for Fe.  Ne and Si have higher peak
abundance values than the other elements, but Ne abundances are skewed
higher than the other elements throughout the remnant.  In most of the
remnant, O, Mg, Si, and Fe abundances are generally between 0.3 to
0.6-0.7 solar, but the Ne abundances fall mostly between 0.55-0.85
solar.  Ne and Mg lines are blended with Fe L emission, so their
abundances are somewhat more model-dependent than those of elements
with cleaner line emission, such as Si.  It should also be noted here
that sub-solar abundances are the norm in Puppis A when the spectra
are fitted with single-component models.  For example, Tamura (1995)
typically required abundances from 0.1 to 0.6 solar for the ASCA
spectra.

The O and Ne abundance maps are patchier than the others.  The fitted
abundances are lowest in the SE Low SB region, and highest in the
southern peak of the Si Knot; we refer to this as the O Knot Center in
Table 3 ($\alpha$=125.67 $\delta$=-42.84).  The absence of this
feature in the O EQW image (not shown here) underlines the difficulty
of identifying O ejecta from the X-ray images alone; the imaging
analysis is particularly difficult for lines like those of O that lie
in crowded spectral regions.  Only by examining the spectra could we
learn that the O abundance is high at the Si Knot.  Ne shows another
relative enhancement in an extended region just south of the broadband
contours that is also echoed more faintly by O.  This region is
clearly visible in the Ne difference image in Figure 3, so we refer to
it as Ne Knot South in Table 3 ($\alpha$=125.66, $\delta$=-43.12).
For Mg, Si and Fe, the Si Knot is the only region with a definite
element enhancement.

We take a closer look at a few spectra selected for their interest, in
Figure 6. The top panel compares the spectrum of a region in the
northern peak of the Si Knot with one in the NE SF.  Both have strong
Si line emission in the Si line image in Figure 2, but only the Si
Knot knot has enhanced EQW (Figure 4) and enhanced Si abundance from
spectral fitting.  The Si He-like blend is indeed stronger at the
knot.  From Table 3, we see that the column densities and temperatures
are similar for the two fits, whereas the NE SF has a somewhat higher
ionization age.  All the fitted element abundances are higher by at
least a factor of two at the Si Knot compared to the NE SF, with Si
being higher by a factor of four.  Given that the statistical errors
are quite tight, the Si Knot is clearly enriched with ejecta.

The second panel shows two spectra with O and Ne enhancements---those
of the O Knot Center (the southern peak of the Si Knot) and the Ne
Knot South.  While the O Knot Center also has higher-than-average
abundances for all the elements except perhaps Fe, the Ne Knot South
does not really show evidence for significant enhancement of elements
other than Ne, which is especially prominent; O is only slightly
higher than average.

In the bottom panel of Figure 6, we compare regions without
significant abundance enhancements---the SE Arm extending south from
the BEK, the SE Low SB region, and the eastern edge of the BEK.  The
latter two are prominent in the O VII/O VIII ratio image, and the SE
Low SB region also appears in the O VIII Ly$\alpha$ EQW image.
Enhanced O VII/O VIII in the SE Low SB region and the BEK is affirmed
by the spectra, as the O VII line is clearly stronger in these regions
than in the SE arm, whereas O VIII emission is comparably strong in
all three regions.  Enhanced O EQW in the SE Low SB region is not
affirmed by the fitted O abundance, however.  The explanation would
appear to be that the ``continuum band'' used to subtract the
continuum contribution (``Cont 1'' at 0.70-0.79 keV from Table 2) is
relatively weaker here (due in part to lower Fe abundance), which
artificially enhances the O EQW.  The spectral parameters given in
Table 3 show that these three regions have similar columns,
temperatures, and abundances.  The primary difference is a lower
ionization age for the SE Low SB region, which is enough to shift the
ion balance in favor of He-like O and to enhance the O VII/O VIII line
intensity ratio.  Compared to other regions in the remnant, the SE Low
SB region also has very low column density and relatively low
temperature, which combines with the low ionization age to enhance O
VII relative to O VIII.

\begin{figure}
\centerline{\includegraphics[scale=0.8]{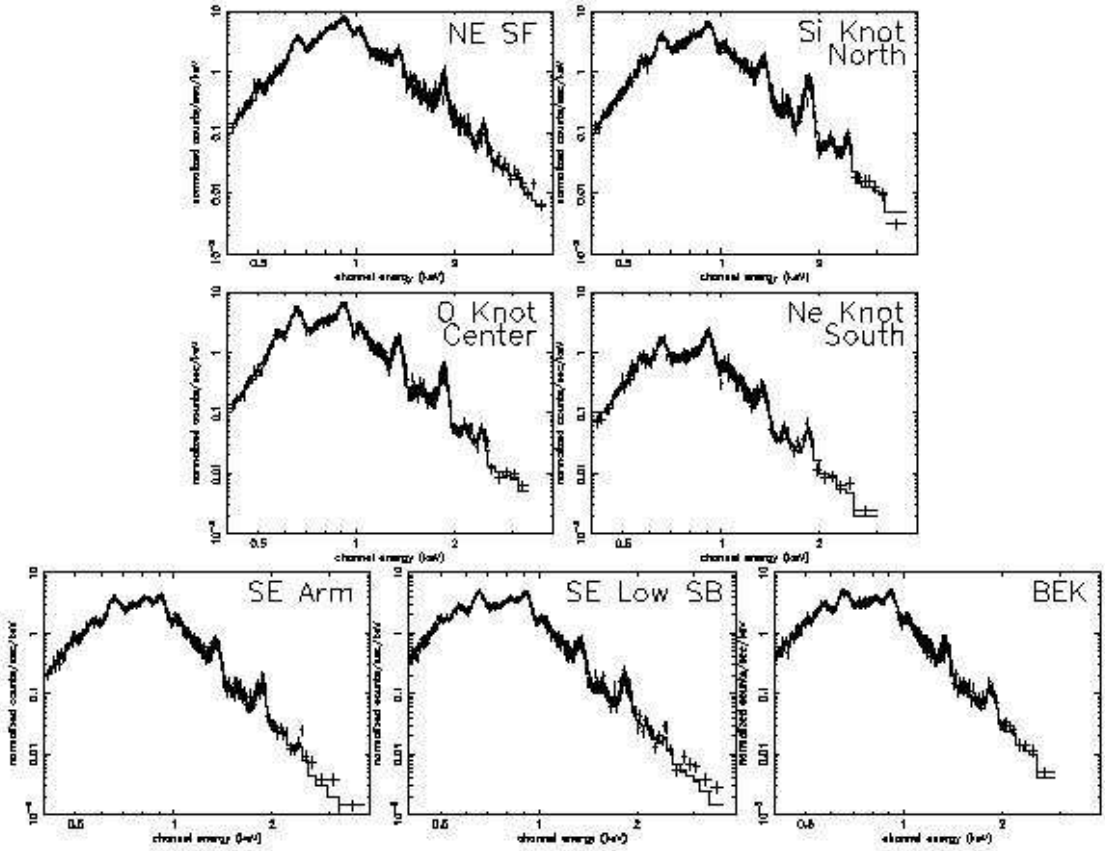}}
\vskip -2.0 in
\caption{(Top) Spectra comparing strong Si line emission at the NE SF
(left) and the northern peak of the Si Knot (right).  (Middle)
Spectrum of the O Knot Center (the southern peak of the Si Knot) (left) and the Ne Knot South (right).
(Bottom) Spectra from SE Arm with low O VII/O VIII and O VIII EQW
(left) and the SE Low SB region with high O VII/O VIII and O VIII EQW
(middle) and the eastern edge of the BEK (right). Fitted parameters
are given in Table 3.}
\end{figure}
\begin{deluxetable}{lllllllll}
\tabletypesize{\scriptsize}
\rotate
\tablecaption{Spectral Fits}
\tablewidth{0pt}
\tablehead{\colhead{Region}&\colhead{$\chi^2, \chi^2/dof$}&\colhead{N$_{\rm H}^*$}&\colhead{kT}&\colhead{$n_et$}&\colhead{O}&\colhead{Ne}&\colhead{Si}&\colhead{Fe}\\
\colhead{J2000 RA, DEC}&\colhead{}&\colhead{$10^{22}$ cm$^{-2}$}&\colhead{(keV)}&\colhead{$10^{11}$cm$^{-3}$s}&\colhead{$\odot$}&\colhead{$\odot$}&\colhead{$\odot$}&\colhead{$\odot$}}
\startdata
NE SF & 587.4, 1.41 & 0.345$\pm0.005$ & 0.708 (0.707-0.727) & 2.63e11 (2.56-2.72e11) & 0.53 (0.51-0.56) & 0.64 (0.63-0.67) & 0.41 (0.38-0.43) & 0.45 (0.44-0.47) \\ 
125.89, -42.79 &&&&&&&&\\
Si Knot & 558.6, 1.56 & 0.33$\pm0.01$ & 0.88 (0.87-0.91) & 6.5e10 (6.3-6.9e10) & 0.86 (0.84-0.89) & 1.27 (1.24-1.33) & 1.5 (1.4-1.6) & 0.94 (0.90-0.97) \\
125.70, -42.76 &&&&&&&&\\
O Knot Center & 510.7, 1.49 & 0.35 (0.33-0.36) & 0.77 (0.73-0.86) &
7.1e10 (6.2-7.2e10) & 1.3 (1.1-1.4) & 1.8 (1.7-2.0) & 1.3 (1.2-1.3) 
& 0.57 (0.54-0.65) \\
125.67, -42.84 &&&&&&&&\\
Ne Knot South & 271.0, 1.14 & 0.30$\pm 0.01$ & 0.66 (0.64-0.68) & 5.30e10 (5.28-5.30e10) & 0.73 (0.70-0.75) & 1.41 (1.35-1.46) & 0.48 (0.39-0.54) & 0.41 (0.38-0.44) \\
125.66, -43.12 &&&&&&&&\\
SE Arm & 479.2, 1.45 & 0.22$\pm0.01$ & 0.50 (0.48-0.52) & 1.95e11 (1.62-2.03e11) & 0.42 (0.41-0.45) & 0.67 (0.63-0.70) & 0.47 (0.43-0.51) & 0.50 (0.45-0.54) \\
126.05, -43.15 &&&&&&&&\\
SE Low SB & 489.8, 1.42 & 0.173 (0.166-0.183) & 0.64 (0.61-0.68) & 8.2e10 (7.4-9.0e10) & 0.35 (0.32-0.37) & 0.61 (0.59-0.63) & 0.43 ( 0.37-0.47) & 0.39 (0.36-0.42) \\
125.91, -43.14 &&&&&&&&\\
BEK, east edge & 363.7, 1.23 & 0.15 (0.14-0.17) & 0.58 (0.52-0.59) & 1.1e11 (1.0-1.3e11) & 0.38 (0.35-0.40) & 0.69 (0.64-0.74) & 0.40 (0.34-0.49) & 0.43 (0.38-0.47) \\
126.08, -42.98 &&&&&&&&\\
\enddata

\tablenotetext{*}{Error ranges given are 90\% confidence for a single interesting parameter.}
\end{deluxetable}

\section{Discussion}

To summarize the observational results, Suzaku clearly detects X-ray
emitting ejecta in Puppis A, and shows where it is
located.  The most striking feature of the spectral maps is the
spatial correlation of the element abundances, especially at the
location of the Si Knot.  All the elements are found to be
particularly enriched this region, with Si appearing to show the most
enrichment relative to other regions.  Ne, and to a lesser extent O (
but not other elements), is also enriched in the south center of the
remnant.

The two peaks of the Si Knot are both regions with somewhat
higher-than-average temperatures, but even a forced temperature of 0.6
keV closer to the remnant average requires clearly enhanced element
abundances (Si=1.3 and O=0.65 for the weaker northern peak of the
knot).  We consider it more likely that the temperature difference is
real and is associated with the different composition of the knot
compared to its surroundings.

The spectral parameters also show strong patterns.  The high column
density to the north (combined with relatively uniform temperature)
echoes the patterns seen in spectral hardness by ROSAT.  There is some
hint that the fitted column densities may be correlated with detector
position in the two northernmost fields (covering the northern peak
and the interior southwest of it), with the center of the detector
giving higher values.  The other fields do not give the same
impression, however, and the large-scale differences in column density
are certainly robust.  The ridge of high ionization age that follows
the boundary of the northeast shock front is qualitatively consistent
with the strong interaction of the blast wave seen in that region.
Radio HI observations show a very large molecular cloud whose straight
edge fits against the X-ray boundary of the remnant (Reynoso et
al. 1995).  The blast wave will be strongly decelerated as it
propagates through the dense cloud, but the reflected shock will be
associated with higher temperature gas (cf. Levenson et al. 2002).
The difference in ionization age is a factor of several on the ridge
compared to the region behind it, and implies a comparable density
difference assuming that the shock times are roughly the same.  Higher
angular resolution observations would be useful to make more reliable
quantitative estimates, but in any case, the relative narrowness of
the ridge seems to suggest that there was a sudden increase in the
ambient gas density, which presumably occured at the boundary of the
molecular cloud.

The Chandra study of the BEK region in Puppis A showed that there are
significant spectral differences even within the BEK.  The eastern
0.5$'$ of the region is a compact knot having different spectral
parameters from the portion of the BEK to the interior. These
differences are unfortunately not accessible to this study, being on
too small an angular scale compared to the angular resolution of the
spectral grid.  Nevertheless, it is encouraging to see that other,
somewhat larger scale spectral features seen in the BEK region by
Chandra are also seen by Suzaku.  For example, there is a clump of
high ionization age gas inside the main part of the BEK which
corresponds well with the bottom end of the ``ridge'' of high
ionization age in the Suzaku map.  The increase in temperature just
north of the BEK is also reproduced by the Suzaku analysis.  The only
sufficiently large angular-scale spectral feature that we did not see
reproduced in the Suzaku spectral images is the higher column density
gas, not only at the eastern edge of the BEK, but also at the eastern
boundary above and below it.  We do not have a complete explanation
for this, but it may be related in part to the contaminants that build
up on CCDs, since these reduce low energy sensitivity and can be
masked as increased absorption column densities; this is relevant for
both the Suzaku and Chandra CCDs.

The overall subsolar element abundances also require some comment.  As
noted earlier, similarly low abundances had been obtained by Tamura
(1995) in his analysis of ASCA X-ray spectra, but there are otherwise
no indications that the interstellar medium around Puppis A has
depleted element abundances.  Dust depletion is a possibility, given
the brightness of the remnant at infrared wavelengths, and the
excellent correlation of infrared and X-ray morphologies (Arendt et
al. 1990).  Dust is clearly present in the remnant and is apparently
collisionally heated.  The signature of dust depletion in the X-ray
emission would be higher Ne abundance to that of O, Mg, Si, and Fe,
since Ne is not depleted onto dust grains as are these other elements.
The Ne abundance does have a tendency to be slightly higher than that
of O in Puppis A, but so is the Si abundance.  Moreover, the Ne
abundance in CCD-resolution spectra is subject to additional
uncertainty because the Ne lines are blended with the complicated Fe L
line emission.  In the south-central Ne knot of Puppis A, the fitted
Ne abundance is near the solar value and significantly enhanced
relative to the other elements; perhaps this is a region where dust
depletion is important.  The Ne abundance is nevertheless generally
fitted at subsolar values, however, so it would appear that the
overall abundances are indeed subsolar, independent of any depletion
of some of the elements onto dust grains.

Element abundances obtained from an overly simple spectral model may
be inaccurate, but there are no definite indications of additional
components in Puppis A that would, if present and unaccounted for,
result in artifically lower element abundances.  Neither is Puppis A
an isolated incidence of low abundances inferred for the ISM-phase of
a supernova remnant.  The Cygnus Loop in particular has yielded
extremely low ISM-phase abundances of 0.1-0.2 solar to analysis using
data from ASCA (Miyata et al. 1994), Chandra (Levenson et al. 2002,
Leahy et al. 2004), XMM-Newton (Tsunemi et al. 2007) and Suzaku
(Miyata et al. 2007).  Finally, the interstellar medium in the 
Magellanic Clouds is known to have subsolar abundances that are in
fact reproduced by Sedov models fitted to the entire remnant spectrum
(Hughes et al. 1998)\footnote{Plane-parallel shocks should provide a
good approximation to Sedov models (Borkowski et al. 2001).}.

\subsection{Lack of Widespread O Enrichment}

The modest O ejecta enrichment seen with Suzaku is at odds with the
significant enrichment of O and Ne relative to Fe that was reported by
Canizares \& Winkler (1981; hereafter CW81).  In most regions of
Puppis A, we do not see O significantly enhanced relative to Fe, even
where O is enhanced above its average value.  This echoes earlier
spectral studies of Puppis A with the Einstein Solid-State
Spectrometer (Szymkowiak 1985) and the ASCA Solid-State Imaging
Spectrometer (Tamura 1995), neither of which found a significant O/Fe
enrichment anywhere in Puppis A.  The FPCS has also reported
enrichments of O relative to Fe for other sources, including
M87 and the supernova remnant N132D.  For M87, XMM-Newton Reflection
Grating Spectrometer observations indicated a O/Fe abundance ratio of
0.5 (Sakelliou et al. 2002), compared to the factor of 3-5 enrichment
reported by the FPCS (Canizares et al. 1982).  For N132D, ASCA finds that
O/Fe is roughly solar (Hughes, Hayashi, \& Koyama 1998), in contrast
to values of twice solar or higher as deduced from Einstein
observations (Hwang et al. 1993).

The Einstein FPCS observations were the first high-resolution X-ray
spectra ever obtained of non-solar cosmic sources---and the FPCS
observation of Puppis A was the very first of these.  The FPCS used a
number of different crystals to scan limited energy ranges around a
feature of interest.  The calibration was thus challenging, and not
fully optimal because the calibration data files taken just prior to
launch were unusable.  The final calibration was based on extensive,
earlier laboratory measurements\footnote{Mark Schattenburg, private
communication}.  For M87, the relative abundance estimate was based on
the O VIII Ly $\alpha$ line at 654 eV and the Fe XVII line at 826 eV,
for which two different crystals were required to scan the separate
lines, adding to the calibration uncertainties.  This uncertainty is
eliminated for the observations of Puppis A and N132D since a single
crystal was used to scan the O VIII Ly $\gamma$ (816 eV) and Fe XVII
lines for the relative abundance determination.  There is added
uncertainty, however, in that this is a crowded spectral region, and
the relatively weak Ly $\gamma$ line had to be modelled carefully.

For Puppis A, the assumptions made by CW81 to interpret their spectra
were reasonable enough at the time, but updated X-ray observations and
atomic physics show some possible pitfalls.  These can be summarized
as arising from (1) the assumption of a single-component (i.e., single
temperature, single ionization age) plasma within the large FPCS
apertures (2) comparison to a solar coronal spectrum with a
temperature distribution that turns out not to be appropriate for most
of Puppis A (3) the assumption that Fe XVII is always the dominant Fe
ion species when O VII and O VIII ions coexist. (1) Spectral maps of
Puppis A, including the one in this work, clearly show a wide variety
of temperatures, column densities, and ionization timescales
throughout the remnant.  (2) From the same maps, we see that
temperatures in Puppis A are relatively high, averaging near 0.6 keV.
By contrast the solar coronal spectrum used to calibrate the FPCS
observations peaked at roughly 0.2 keV (3) Evaluation of the ion
populations using the NEI models in XSPEC show that O VII, O VIII, and
Fe XVII do not necessarily coexist as assumed, particularly at
temperatures below kT=0.2 keV and, more pertinently for this
discussion, at temperatures above kT=0.6-0.7 keV.  The same evaluation
of ion populations show that they can vary by factors of at least 3-5
in the optically thin NEI models, particularly when ionization effects
are also taken into account.

Given the foregoing, we believe that the current Suzaku abundance
results are more accurate, the high spectral resolution of the FPCS
notwithstanding.

\subsection{Spatial Correlations}

Although the peak abundances for the different elements vary from
subsolar (Fe) to nearly twice solar (Ne and Si), the abundance maps
all have the same general appearance, with the double-peaked Si Knot
being the most prominent feature throughout.  At the southern peak,
all the elements with line emission have element abundances that are
significantly higher than average, and reach their highest values (in
this survey).  The northern peak of this knot is relatively weaker for
O, Ne, and Mg compared to Si, but when the errors on the fitted
abundances are considered (Table 3), there is no significant
difference in the Si/O abundance ratio at the northern vs the southern
peak.  The presence of Si ejecta at a radius comparable to that of O
ejecta is certainly interesting, because supernova ejecta may be
either hydrostatic or explosive in origin; O is synthesized completely
hydrostatically by the progenitor before the explosion, and Si in the
inner layers of the exploding star.  Thus Si originates close to the
center of the explosion, but has been propelled outward in the
northern region of Puppis A and mixed with O and Ne ejecta.

Most of the fast optically emitting ejecta [O III] knots reported by
Winkler et al. (1985) are in the eastern half of Puppis A observed
with Suzaku.  The bulk of them lie 25 to 90 degrees east of north, at
5$'$ to 9$'$ from the explosion center determined from their proper
motions (this center is indicated in the Si map subpanel of Figure 5).
Many of the optical knots are part of a large (roughly $8'\times 4'$)
tangle of [O III] filaments lying to the south and east of the
southern peak of the X-ray Si Knot.  A smaller clump of optical knots
touches the X-ray knot at its southeastern boundary.  Thus, while the
optically emitting O ejecta are in the general vinicity of the O
element abundances identified through the X-ray spectral fits, there
is no close physical correspondence between the optical and X-ray
emitting ejecta.  On the whole, the X-ray and optically emitting O
ejecta appear to be spatially disjoint.


It is clear from the Suzaku observations that the Si knot is the only
example of significant Si ejecta enhancement in the eastern half of
Puppis A.  In a result simultaneous with the preparation of this
paper, Katsuda et al. 2007 (in preparation) independently confirm the
Si enhancement here using XMM-Newton data, which provides
complementary coverage to Suzaku of the western half of the remnant.
They survey a larger fraction of the remnant, but do not find other
regions with substantial Si ejecta enrichment.  If we allow that the
average interstellar element abundances near Puppis A are below the
solar value, it would appear that Fe is enhanced at the Si Knot as
well, since the fitted Fe abundances are higher here than elsewhere in
the remnant.  The Si Knot is thus the only recognizable structure that
contains significant amounts of any explosively synthesized ejecta,
whether Si or Fe.  Unlike the case in Cas A, where there is evidence
for physical separation of the Si and Fe ejecta (Hughes et al. 2000),
Puppis shows Si and Fe ejecta that appear to be coincident with each
other, as well as being mixed with lower Z elements.

\subsection{Knot Ejecta Mass Ratios}

The north central knot being the cleanest sample of ejecta in the
remnant, we consider how its element abundances compare to the
predictions of nucleosynthesis models.  We convert the fitted
element abundances in Table 3, which are by number relative to the
solar values of Anders \& Grevesse (1989), to the ratio of the element
mass relative to that of Si.  The observed mass ratios are compared to
various calculations of core-collapse supernova nucleosynthesis by
Woosley \& Weaver (1994), Thielemann et al. (1996), Rauscher et
al. (2002) and Limongi \& Chieffi (2003).

Nucleosynthesis calculations are complex, and are sensitive to the
treatment of the stellar and nuclear physics, some of which is not yet
fully understood.  For these reasons, models have yet to converge in
detail on what the element mass ratios should be for a progenitor of
given mass.  The models we consider predict that a Type II explosion
for a 25 M$_\odot$ progenitor should produce 0.1-0.3 M$_\odot$ Si
ejecta, with the models of Thielemann et al. giving Si masses at the
low end of this range.  Considering the various model calculations for
abundances of O, Ne, Mg, and Fe relative to Si, we do not find perfect
agreement with the observed mass ratios, but progenitors between 15
and 25 M$_\odot$ appear to be generally indicated.  In no case do we
find a plausible model for a progenitor more massive than 30
M$_\odot$, whereas models are seldom computed for masses below about
15 M$_\odot$.

Of course, this particular knot need not even be a representative
sample of the global ejecta abundances, but at the least it is
reassuring that the high abundances of O, Ne and Mg relative to Si
require a core-collapse rather than a thermonuclear explosion.

It is challenging to put tight constraints on the progenitor of Puppis
A from the X-ray observations because the remnant is relatively old
and most of the X-ray emission is dominated by interstellar material,
some of which is highly structured as dense clouds.  Even where ejecta
are present, they are mixed with ISM, but complex models cannot always
be reliably constrained using moderate-resolution X-ray spectra.  We
have only considered simple one-component models.  Technically, the
presence of a second component could skew the derived ejecta, but this
should tend to increase the element abundances.  We found that if we
did attempt to add a second thermal component representing emission
behind the blast wave (hence with the blast wave abundances fixed at
0.4 solar), the fitted ejecta abundances increased overall by a factor
of two or so; the temperature and ionization age for the Si Knot did
not change much, increasing by no more than 10-20\%.  We chose not to
use two-component fits throughout because the spectra generally cut
off above 2-3 keV, and we considered it difficult to put believable
constraints on a second component.  In any case, this is a
conservative assumption for identifying ejecta enhancements, as we
have pointed out.

In spite of the remnant's advanced age, the ejecta in this mature
remnant are tantalizing to pursue for the hints it may give to the
explosion that formed it.  Ideally, better angular resolution is 
desirable since Chandra observations (Hwang et al. 2005) have hinted
at relatively small localized regions with abundance enhancements.
These may contribute to the patchy, low-level enhancements seen with
Suzaku.  Aside from the X-ray emitting Si Knot, the remaining Si may
either already be shocked and cooled, shocked and diluated, or still
unshocked.  If the former, we might hope to see it in optical
emission, but optical [S II] emission in Puppis A resembles H$\alpha$
emission rather than the X-ray Si emission, and appears more likely to
be associated with blast wave emission (P. F. Winkler, private
communication).  Shocked and diluted Si will be difficult to detect
with the significant dilution by the ISM expected for Puppis A after
4000 years.  Roughly 0.1 M$_\odot$ of Si is present in a few hundred
solar masses of interstellar material at 0.4 solar abundances.
Unshocked ejecta might be revealed by infrared observations, but the
available data are unfortunately insufficient for identifying ejecta
(Arendt et al. 1990).  We hope that future observations will shed
interesting light on the ejecta in Puppis A.


\acknowledgments 

We have benefitted from scientific discussions with Satoru Katsuda and
other collaborators on the XMM-Newton observation, and from a careful
reading by an anonymous referee.  We are grateful to Hideyuki Mori for
expert help in configuring the observations, Mark Schattenburg, Dale
Graessle, and Claude Canizares for discussion of the FPCS observations
and calibration, Takashi Okajima for discussion of the Suzaku
mirrors, and Frank Winkler for sharing his optical images.

\end{document}